\documentclass[conference]{IEEEtran}
\IEEEoverridecommandlockouts

\usepackage{cite}
\usepackage{amsmath}
\usepackage{amssymb}
\usepackage{gensymb}
\usepackage{xcolor}
\usepackage{url}
\usepackage{graphicx}
\usepackage[ruled,vlined]{algorithm2e}
\usepackage{multirow}

\DeclareMathOperator*{\argmin}{arg\,min}
\def\ut{U_n(t)}
\def\to{\tau_n^o}
\def\te{\tau_n^e}
\def\df{f_s}
\def\dt{\Delta \tau}
\def\sinc{\text{sinc}}

\begin{document}

\title{A Secure Dual-Function Radar Communication System via Time-Modulated Arrays\\\thanks{Work supported by NSF under grant ECCS-2033433.}}

\author{Zhaoyi Xu and Athina P. Petropulu \\
	Department of Electrical and Computer Engineering, Rutgers University, Piscataway, NJ 08854 \\
	E-mail:  \texttt{\{zhaoyi.xu,athinap\}@rutgers.edu}}

\maketitle

\begin{abstract}
Dual-function radar-communication (DFRC) systems offer high spectral, hardware and power efficiency, as such are prime candidates for  6G wireless systems. DFRC systems use the same waveform for simultaneously probing the surroundings and communicating with other equipment. By exposing the communication information to potential targets, DFRC systems are vulnerable to eavesdropping. In this work, we propose to mitigate the problem by leveraging directional modulation (DM) enabled by a time-modulated array (TMA) that transmits OFDM waveforms. 
DM can scramble the signal in all directions except the directions of the legitimate user. However, the signal reflected by the targets is also scrambled, thus complicating the extraction of target parameters.
We propose a novel, low-complexity target estimation method that estimates the target parameters based on the scrambled received symbols.
We also propose a novel method to refine the obtained target estimates at the cost of increased complexity. With the proposed refinement algorithm, the proposed DFRC system can securely communicate with users while having high-precision sensing functionality.
\end{abstract}

\vspace{-1mm}
\section{Introduction}
\vspace{-1mm}
In today’s technology, radio frequency (RF) front-end architectures are very similar in radar and wireless communication systems. Further, by constantly seeking more bandwidth, wireless communication systems have been using frequency bands that used to be occupied by radar.
For those reasons, there has been a lot of interest in co-designing spatially distributed radar and communication devices that can coexist in the spectrum by controlling the interference \cite{Li2016SpectrumSharing}, or designing systems that integrate communication and sensing function in one platform. The latter systems can be implemented in low-weight, low-cost devices that can efficiently use power and the radio spectrum for both communication and sensing.
{Dual Function Radar-Communication (DFRC) systems} \cite{dfrcs, Hassanien16,Fan2020, Ma20} is a class of integrated radar-communication  systems that use  the  same waveform as well as the same hardware platform for both sensing and communication purposes. As a result,  DRFC systems offer the highest spectral efficiency among integrated systems and their transmitter hardware is a lot simpler than typical integrated systems. 
These benefits  make  DFRC technology very desirable for vehicular networks, WLAN indoor positioning, and unmanned aerial vehicle networks \cite{adas,jrc,5gvv}. 


Physical layer security (PLS) in wireless communication systems has been widely investigated and developed for decades. 
The initial idea was to design the system so that the   signal-to-interference-plus-noise (SINR) ratio difference between  eavesdroppers and legitimate users, also referred to as  secrecy rate, is maximized. This can be achieved via proper  precoder design  \cite{Dong2010PHY,Li2011ergodic,Li2011ergodic2,Li2011cooperative},
or transmission of  artificial noise (AN)\cite{zhang2019AN,wang2017AN} to confound the eavesdroppers.
However, the aforementioned methods  require prior knowledge of eavesdropper channel state information (CSI), or statistics of the eavesdropper CSI, while  AN may give rise to interference to the CUs. 

Directional modification (DM) is a newer direction in PLS, that has drawn a lot of attention in recent years. DM modulates the antenna transmissions so that the communication information is distorted in all directions except the directions of the communication users (CU). Thus, DM makes it difficult for an eavesdropper who is located in a different direction than the legitimate users  to  interpret the communication  signal it receives \cite{daly2009directional}.
Generally, DM can be achieved by appropriately designing  the antenna  weights~\cite{Daly2009DM,Ding2015DM,Hafez2015DM}, or via symbol-level precoding~\cite{Alodeh2016DM,Alodeh2015constructive,Masouros2015exploiting} that creates constructive interference between the transmitted data symbols.
DM does not rely on eavesdropper CSI, and causes no interference to the CUs. Another way to implement DM is via time-modulated arrays (TMA), i.e.,
arrays in which the antennas connect and disconnect  to the radio-frequency (RF) chains during the transmission interval in a controlled manner. 
In \cite{Ding2019time_modulated}, an OFDM DM transmitter using TMA is proposed for achieving PLS in a wireless communication system.
Using a periodic connect/disconnect pattern for each transmit antenna with the same period as the OFDM signal gives rise to harmonic signals at the subcarrier frequencies \cite{Ding2019time_modulated}. Due to these harmonic signals, the data symbols of each subcarrier are scrambled by data symbols from all other subcarriers. By appropriately designing the activation patterns and also by applying phase steering weights on each antenna, the data symbols transmitted along the desired direction remain in tack,  while the data symbols transmitted along other directions are scrambled.
%

In this paper, we propose a novel way to design a secure DFRC system based on an  OFDM DM transmitter as that of  \cite{Ding2019time_modulated}.
While the communication component of the DFRC is the same as that of \cite{Ding2019time_modulated}, here, we focus on the radar component.
Assuming that the radar targets are not in the same direction as the CUs,  the signal reflected back to the DFRC receiver contains distorted data symbols. The coupling between scrambled data symbols from all subcarriers and radar target parameters prevents the direct estimation of radar targets. We propose a novel approach to extract radar target parameters based on the scrambled radar returns. The proposed method first coarsely estimates the target angles and the scrambled data symbols and then estimates the target parameters via discrete Fourier transform (DFT) operations. The coarse estimation has low complexity but limited resolution. We also propose a novel algorithm to improve resolution at the expense of increased complexity.

Time modulation enabled DM has also been investigated in the context of DFRC systems \cite{TMA_DFRC2022,TMA_DFRC2018}. However, \cite{TMA_DFRC2022,TMA_DFRC2018} only consider single-carrier waveforms, which achieve low bit rate. 
In this paper,  we consider DRFC systems using OFDM waveforms, which enable much higher bit rate. Also due to the multiple subcarriers and the scrambled data symbols, the sensing task is more complicated.

\vspace{-1mm}
\section{OFDM System with Time-Modulated Array}
\vspace{-1mm}
Let us consider a TMA comprising a  uniform linear array  with $N_t$ antennas, spaced by $d_t$, connected to one RF chain.

The system transmits OFDM signals with carrier frequency $f_c$ and subcarrier spacing $\df$. There are in total $N_s$ subcarriers.
The binary source data are divided into $N_s$ parallel streams,  are modulated via phase-shift keying (PSK) modulation, or quadrature amplitude modulation (QAM), and are distributed to the $N_s$ OFDM subcarriers.
Each antenna is assigned the same data symbols. 
Each antenna applies an inverse discrete Fourier transform (IDFT) on the data symbols, pre-appends a cyclic prefix (CP), applies a phase weight, converts the samples into an analog multicarrier signal and transmits it with carrier
frequency $f_c$; this signal  will be referred to as OFDM symbol, and has duration $T_p$. Note that in an OFDM system, $T_p = 1/\df + T_{cp}$ where $T_{cp}$ is the duration of the CP.

The antennas connect and disconnect to  the RF chain  via  switches, so that during each channel use, each antenna stays connected during a specified time interval.
Let $U_n(t)$ be a periodic square waveform with period $T_p$, controlling the time modulation pattern of the $n$-th antenna. 
%
The signal transmitted towards direction $\theta$ can be written as
\begin{align}
    x(t,\theta) 
    & = \sum_{n=0}^{N_t-1} e^{-j2\pi nd_t \frac{\sin{\theta}}{\lambda}}w_n U_n(t) e(t).
    \label{eq:signal1}
\end{align}
where $\lambda$ is the wavelength of the carrier frequency and $w_n$ is the weight for the $n$-th antenna, and
\begin{equation}
    e(t) = \sum_{\mu = 0}^{N_p-1}\sum_{s=0}^{N_s-1}d(s,\mu) e^{j2\pi (f_c+s\df)t} rect(\frac{t-\mu T_p}{T_p}),
\end{equation}
where $N_p$ is the total number of OFDM symbols, $d(s,\mu)$ is the data symbol transmitted on the $s$-th subcarrier during the $\mu$-th OFDM symbol and $rect(t/T_p)$ denotes a rectangular pulse of duration $T_p$. {The antenna weights are set as  $w_n = e^{j2\pi nd_t \sin{\theta}/c}$, causing the transmitted signal from all antennas to add up coherently in direction $\theta$.}

In this paper, we consider using single-pole-single-throw (SPST) switches, i.e., the switches have two states ``on" and ``off". 
Let $t_n^o$ and $t_n^e$ respectively denote the ``turn-on" and ``turn-off" time of the $n$-th antenna. For  simplicity, we use  normalized on and off time, $\to = t_n^o\df  \in [0,1]$ and $\te = t_n^e\df \in [0,1]$.
Let $\Delta\tau_n = \te - \to$ be the normalized duration of the duty cycle of the $n$-th antenna.
Let us take $\to< \te$,  and set $U_n(t)$ to be $1$ only when $t/T_p\in[\to, \te]$, otherwise it is $0$.
Based on this activation pattern,  the periodic square waveform $U_n(t)$  can be expressed as Fourier series, i.e., 
\begin{multline}
    \ut 
    = \sum_{m=-\infty}^{\infty}e^{j2\pi m\df t} \dt_n\sinc(m\pi\dt_n)\times\\ e^{-jm\pi(2\to+\dt_n)}
    \label{eq:ut}
\end{multline}
where $\sinc(\cdot)$ denotes the sinc function. 

Provided that the data symbols are independent between OFDM symbols, in the following analysis we will only focus on one OFDM symbol, and
discard the rectangular function.
By substituting \eqref{eq:ut} into \eqref{eq:signal1}, 
\textcolor{black}{and focusing on the signal within the frequency range $[f_c, f_c+N_s\df]$, }
we get
\begin{equation}
    \begin{aligned}
    x(t,\theta) 
    = & \sum_{s=0}^{N_s-1}d(s,\mu) e^{j2\pi (f_c+s\df)t}\\&\times\sum_{m=-s}^{N_s-s-1}  e^{j2\pi m\df t}V(m,\to,\dt_n,\theta),
    \\ = & \sum_{s=0}^{N_s-1}\sum_{m=-s}^{N_s-s-1}d(s,\mu) e^{j2\pi [f_c+(s+m)\df]t}V(m,\to,\dt_n,\theta),
    \label{eq:transmitted signal}
\end{aligned}
\end{equation}
where
\begin{equation}
    \begin{aligned}
    V(m,\to,\dt_n,\theta) = &\sum_{n=0}^{N_t-1} e^{-j2\pi nd_t \frac{\sin{\theta}}{\lambda}}w_n\dt_n\\
     &\times \sinc(m\pi\dt_n)e^{-jm\pi(2\to+\dt_n)}
    \label{eq:V}
    \end{aligned}
\end{equation}
contains the Fourier series coefficients of both the fundamental signal (when $m=0$) and the harmonic signals (when $m \neq 0$) due to the time modulation, the phase delay due to angle, and the antenna weight.

Let $i = s+m$, \eqref{eq:transmitted signal} can be recast as
\begin{equation}
    \begin{aligned}
    x(t,\theta) = & \sum_{i=0}^{N_s-1}\sum_{s=0}^{N_s-1}d(s,\mu)V(i-s,\to,\dt_n,\theta)e^{j2\pi i\df t},
    \end{aligned}
\end{equation}
when it can be seen that the data symbol on subcarrier $i$ is the summation of data symbols modulated by the corresponding coefficients $V(i-s,\to,\dt_n,\theta)$ from all subcarriers. 

By properly designing the ``turn-on" time instant $\to$ and the duty cycle for each antenna $\dt_n$ \cite{Ding2019time_modulated}, the following conditions can be satisfied:
\begin{equation}
    \begin{cases}
        V(m = 0,\to,\dt_n,\theta=\theta_0) \neq 0,\\
        V(m \neq 0,\to,\dt_n,\theta=\theta_0) = 0,\\
        V(m \neq 0,\to,\dt_n,\theta \neq \theta_0) \neq 0. 
    \end{cases}
    \label{eq: condition1}
\end{equation}
At the direction $\theta_0$, only the fundamental signal will be preserved while at other directions the received signal contains both the fundamental signal and the harmonics. Thus at direction $\theta \neq \theta_0$, the received data symbol will be scrambled with the data symbols from all other subcarriers. Such a transmitter could provide an enhanced security against eavesdroppers without prior knowledge on their locations and the corresponding channels. 
Note that the benefits of the TMA come at a cost of reduced transmitted power as the duty cycle $\dt_n$ is less than $1$.

\vspace{-1mm}
\section{Radar Sensing with Scrambled Symbols}
\vspace{-1mm}
In this section, we  propose a novel target estimation method which first estimates the scrambled data symbols and then extracts the target parameters. Let us consider a receive array with $N_r$ antennas spaced by $d_r$.

Let the CU be at direction $\theta_c$, where the data symbols are preserved.
Assuming that the radar targets are  not in the same direction as the CU, the OFDM signals that arrive at them are scrambled by the data symbols from all subcarriers as shown in \eqref{eq:transmitted signal}. 


Let us assume that there are $K$ targets and characterize them by the corresponding angle $\theta_k$, range $R_k$ and radial velocity $v_k$. The baseband signal received by the $m$-th antenna during the $mu$-th OFDM symbol is
\begin{align}
    y(m,t) = &\sum_{k=0}^{K-1}\beta_k e^{-j2\pi md_r\frac{\sin{\theta_k}}{\lambda}} \sum_{s=0}^{N_s-1}d'(s,\mu,\theta_k)e^{-j2\pi s\df \frac{2R_k}{c}} \nonumber \\
    &\times  e^{j2\pi f_d(k,s)t} e^{j2\pi (s\df)t} + n(t),
\end{align}
where $\beta_k$ is a complex coefficient accounting for the scattering process associated with the $k$-th target; $f_d(k,s) = 2v_k(f_c + s\df)/c$ is the Doppler frequency of the $k$-th target on the $s$-th subcarrier with $c$ denoting the speed of light and $n(t)$ denotes independent and identically distributed (i.i.d.) white Gaussian noise with zero mean and variance $\sigma_n^2$; and
\begin{equation}
    d'(s,\mu,\theta) = \sum_{i=0}^{N_s-1}d(i,\mu) V(s-i, \to,\dt_n,\theta).
    \label{eq: scrambled symbol}
\end{equation}
Here, $d'(s,\mu,\theta)$ can be viewed as the scrambled symbol.
As one can see, the scrambling depends on the direction $\theta$.

We assume that the observation time is very small such that the range $R_k$ and Doppler frequency $f_d(k,s)$ remain constant during the observation. 

After sampling the baseband received signal with sampling frequency $F_s = N_s\df$ over the duration $T_p-T_{cp}$, an $N_s$-point discrete Fourier transform (DFT) is applied on the samples to obtain the symbols. The symbol received by the $m$-th antenna on the $i$-th subcarrier during the $\mu$-th OFDM symbol equals
\begin{align}
    d_r(m,s,\mu) = &\sum_{k=0}^{K-1}\beta_k d'(s,\mu,\theta_k)e^{-j2\pi md_r\frac{\sin{\theta_k}}{\lambda}} e^{-j2\pi s\df \frac{2R_k}{c}} \nonumber \\
    &\times  e^{j2\pi \mu T_p f_d(k,s)} + n_s,
    \label{eq: received symbol}
\end{align}
where $n_s$ is the noise on the subcarrier $s$.

In \eqref{eq: received symbol}, the scrambled symbols are coupled with the target parameters, thus one cannot use the conventional symbol-based estimation method in \cite{ofdm} to estimate the target parameters. 

\subsection{Angle Estimation}
Ignoring the noise, \eqref{eq: received symbol} can be viewed as
\begin{equation}
    d_r(m,s,\mu) = \sum_{k=0}^{K-1} A(k,s,\mu)e^{j2\pi \omega(k,s) m},
\end{equation}
where $
    A(k,s,\mu) = \beta_k d'(s,\mu,\theta_k)e^{-j2\pi s\df \frac{2R_k}{c}}e^{j2\pi \mu T_p f_d(k,s)},
$
and $
    \omega(k,s) = -d_r\frac{\sin{\theta_k}}{\lambda}.
$
Assuming that $N_r > K$, and for a fixed $s$, $\{ d_r(m,s,\mu), m=0,...,N_r-1\}$ can be viewed as a sum of $K$ complex sinusoids with frequencies $\omega(k,s)$ and magnitudes $A(k,s,\mu)$. Thus, upon applying an $N_r$-point on that sequence, we can get peaks at frequencies $\omega(k,s)$. Once the frequencies are estimated, the target angles can be computed as
\begin{equation}
    \theta_k = \text{arcsin}\bigg(-\frac{\omega(k,s)\lambda}{d_r}\bigg).
    \label{angle}
\end{equation}

The angle estimation scheme can be repeated on all subcarriers.
Thus, by exploiting frequency diversity, one can find all
occupied angle bins with high probability.

The resolution of the peaks in the aforementioned DFT  depends on the number of receive antennas, $N_r$. We will refer to the angle estimates of \eqref{angle} as \textit{coarse}, and in Sec.~\ref{sec: refine}, we will show how these angles estimates can be refined.

We should note that in the above estimation, the complex amplitude $\beta_k$ does not need to be known, because the coarse angle estimation is along the receive array domain.

\subsection{Range and Velocity Estimation}
\vspace{-1mm}
Once the angle DFT  separates the radar targets into different angle bins, focusing on each  angle bin, we can estimate how the data symbols are scrambled based on \eqref{eq: scrambled symbol}. 
Based on the assumption that the observation time is very small,
the amplitudes of the angle DFT peaks can be written as
\begin{align}
    A(k,s,\mu) &= \beta_k d'(s,\mu,\theta_k)e^{-j2\pi s\df \frac{2(R_k- \mu T_p v_k)}{c}}e^{j2\pi \mu T_p \frac{2v_kf_c}{c}}\nonumber\\
    & \approx \beta_k d'(s,\mu,\theta_k)e^{-j2\pi s\df \frac{2R_k}{c}}e^{j2\pi \mu T_p \frac{2v_kf_c}{c}}.
\end{align}
for $\mu = 0,1,\dots, N_p$.
Here the coefficient $A(k,s,\mu)$ contains the known scrambled data symbols and the phase shifts due to range and velocity.
The range phase shift $e^{-j2\pi s\df \frac{2R_k}{c}}$ only depends on the subcarrier index, $s$, while the velocity phase shift only depends on the OFDM symbol index, $\mu$. As the subcarrier domain and the OFDM symbol domain are orthogonal, we can independently estimate the targets' ranges and velocities within each occupied angle bin.

There can be multiple targets in the same angular bin. Suppose that there are $N_q$ targets at angle $\theta_{k}$.
After the element-wise division on the estimated scrambled symbols at $\theta_k$, the coefficient can be written as
\begin{align}
    A'(k,s,\mu) &= \frac{A(k,s,\mu)}{d'(s,\mu,\theta_k)}\nonumber,\\
    & = \sum_{q = 0}^{N_q-1}\beta_{kq} e^{-j2\pi s\df \frac{2R_{kq}}{c}}e^{j2\pi \mu T_p \frac{2v_{kq}f_c}{c}},
    \label{eq: element-wise division}
\end{align}
where $\beta_{kq}$, $R_{kq}$ and $v_{kq}$ are respectively the coefficient, range and radial velocity of the $q$-th target at angle $\theta_k$.


\eqref{eq: element-wise division} provides an expression that contains ranges and velocities only, while the scrambled data symbols have been eliminated. 
The ranges can then be estimated based on the peaks of an $N_s$-point inverse discrete Fourier transform (IDFT) of $A'(k,s,\mu)$, taken along the $s$ dimension, i.e., 
\begin{align}
    r(k,l,\mu) &= IDFT[A'(k,s,\mu)] =
    \frac{1}{N_s}\sum_{s=0}^{N_s-1}A'(k,s,\mu)e^{j\frac{2\pi}{N_s}sl}, \nonumber \\
    &=\sum_{q=0}^{N_q-1}\frac{\beta_{kq}e^{j2\pi \mu T_p \frac{2v_{kq}f_c}{c}}}{N_s}
    \sum_{s=0}^{N_s-1}e^{-j2\pi s\df \frac{2R_{kq}}{c}}
    e^{j\frac{2\pi}{N_s}sl},
    \label{range}
\end{align}
for $l=0,...,N_s-1$.
The peaks of $r(k,l,\mu)$ will appear  at  positions
 \begin{equation}
    l_{kq} = \Big\lfloor\frac{2N_sR_{kq}\df }{c}\Big\rfloor, 
    \label{rangeind}
 \end{equation}
where $\lfloor\cdot\rfloor$ denotes the floor function.

Then, by performing an $N_p$-point DFT on the range peaks from previous estimation along the dimension $\mu$ , we get peaks at
\begin{equation}
    p_{kq} = \Big\lfloor\frac{2v_{kq}f_cN_pT_p}{c}\Big\rfloor.
    \label{veloind}
\end{equation}
Based on the location of those peaks we can estimate the targets' velocities.

Note that, by performing range estimation within each angle bin, the range estimates are paired with estimated angle. Also, the velocity estimates can be paired with the angle-range estimation by carrying out the DFT on the range IDFT peaks along the OFDM symbol domain. The coefficients $\beta$ are not needed to be known during the estimation, as the range and velocity estimation are respectively along the subcarrier domain and OFDM symbol domain.

\vspace{-1mm}
\section{Target Estimates Refinement}
\label{sec: refine}
\vspace{-1mm}
In the above described estimation process, the angle resolution is limited by the aperture of the physical array, i.e.,  $(N_r-1)d_r$.
Low angle resolution may render some target angles unresolvable. 
In this section, we propose a novel algorithm to refine the target angle and range estimation. Since the angle and range phase shifts are independent of the OFDM symbol index,  the refinement only requires one OFDM symbol. 

Suppose that $N_q$ targets are within the same angle bins but have different angles.
Since the data symbols are scrambled in different ways in different directions, high-resolution angle estimation is required to precisely estimate the different scrambled data symbols inside the same angle bin.
Multiple signal classification (MUSIC) algorithm as a widely used super-resolution method can be used to provide such high-resolution angle estimates. Let the refined angle estimates are $\hat{\Theta} = [\hat{\theta}_0,\hat{\theta}_1,\dots, \hat{\theta}_{N_q-1}]^T$.

Although the angle estimates are refined by the MUSIC algorithm, they cannot be directly used in \eqref{eq: received symbol} due to the coupling between symbols and target parameters. To refine the range estimations and further pair them with the refined angle estimates, we propose to leverage the previously estimated ranges in \eqref{rangeind} and formulate a least-squares optimization problem. For simplicity, we will first start with the case where the targets are at the same range, and then we will discuss the case where the targets are at different ranges.

Let $\mathbf{R} = \{R_0, R_1,\dots, R_{N_q-1}\} $ contains the discretized range space around the already estimated range. The least-squares optimization problem during the $\mu$-th OFDM symbol can be expressed as
\begin{equation}
    \hat{{R}} = \argmin_{R\in \mathbf{R}} \sum_{m=0}^{N_r-1}\sum_{s= 0 }^{N_s-1} |d_r(m,s,\mu) - d'_r(m,s,\mu,R)|^2,
    \label{eq: MLE}
\end{equation}
where 
\begin{align*}
    d'_r(m,s,\mu,R) = \sum_{q=0}^{N_q-1} d'(s,\mu,\hat{\theta}_q)e^{-j2\pi md_r\frac{\sin{\hat{\theta}_q}}{\lambda}} e^{-j2\pi s\df \frac{2R}{c}}
    \label{eq:constructed symbols}
\end{align*}
is the estimated received data symbols. 

When the targets are at different ranges, 
one needs to use all possible combinations of discretized ranges and the refined angle estimates in \eqref{eq: MLE} to refine the range estimates and correctly pair them with the refined angle estimates. 

After the angle-range refinement, the velocity estimation refinement can be also be conducted in the similar way
\begin{equation}
    \hat{{v}} = \argmin_{v\in \mathbf{V}} \sum_{m=0}^{N_r-1}\sum_{s= 0 }^{N_s-1}\sum_{\mu=0}^{N_p-1} |d_r(m,s,\mu) - d'_r(m,s,\mu,v)|^2,
    \label{eq: MLE2}
\end{equation}
where 
\begin{align*}
    &d'_r(m,s,\mu,v) = \nonumber\\&\sum_{q=0}^{N_q-1} d'(s,\mu,\hat{\theta}_q)e^{-j2\pi md_r\frac{\sin{\hat{\theta}_q}}{\lambda}} e^{-j2\pi s\df \frac{2\hat{R}}{c}}e^{-j2\pi \mu T_p \frac{2vf_c}{c} }
    \label{eq:constructed symbols2}
\end{align*}
is based on the refined angle range pairs from \eqref{eq: MLE}.
However, as the velocity needs to be estimated along the OFDM symbol domain, it requires one to estimate the scrambled data symbols in all OFDM symbols, which  introduces  significant computation complexity. To reduce complexity one could formulate the least-squares optimization problems in \eqref{eq: MLE} and \eqref{eq: MLE2}  using a subset of the receive antennas and/or a subset of the subcarriers, however, the reduced numbers of samples would degrade the refinement performance.


Note that the above estimation refinement is for targets within the same angle bin. Thus, if there are $K$ occupied angle bins, we need to repeat the above refinement $K$ times to refine the estimation for all targets. 

The total estimation process is described in Algorithm~\ref{alg:angle-range}. In the first step, the target angles are coarsely estimated via DFT operations along the receive array domain on all subcarriers (see \eqref{angle}). Then based on the coefficients of the angle DFT peaks, the scrambled data symbols can be estimated and eliminated from the angle DFT peak coefficients (see \eqref{eq: element-wise division}). After the element-wise division, range and velocity can be independently estimated via IDFT and DFT operations along the subcarrier domain and OFDM symbol domain respectively. Then the MUSIC algorithm is used to refine the angle estimates and a least-squares optimization problem is formulated based on the refined angle estimates and the previously estimated ranges. By searching over all possible combinations, the results of the least-squares optimization problem are the paired refined angle-range estimates $(\hat{\Theta}, \hat{\mathbf{R}})$.
The velocity estimates can also be refined in a similar way, but at a cost of a significant computation overhead.

\vspace{-3.5mm}
\begin{algorithm}[h]
\SetAlgoVlined
\DontPrintSemicolon
Step 1: Obtain coarse angle estimates via \eqref{angle} on all subcarriers,\;
Step 2: Obtain coarse range estimates via \eqref{rangeind} within the occupied angle bins found in the previous step.\;
Step 3: Obtain refined angle estimates ${\hat{\Theta}}$ using the MUSIC algorithm,\;
Step 4: Formulate and solve a least-squares optimization problem based on the coarse range estimates of Step 2 and the refined angle estimates ${\hat{\Theta}}$ of Step 3 via \eqref{eq: MLE} to obtain refined angle-range estimates $(\hat{\Theta}, \hat{\mathbf{R}})$,\;
Step 5: Formulate and solve another least-squares optimization problem based on the previously refined angle-range estimates $(\hat{\Theta}, \hat{\mathbf{R}})$ to refine the velocity estimates and obtain the final estimates $(\hat{\Theta}, \hat{\mathbf{R}}, \hat{\mathbf{V}})$.
\caption{Target Parameter Estimation Algorithm For a TMA OFDM DFRC system.}
\label{alg:angle-range}
\end{algorithm}
\vspace{-4mm}

In this algorithm, the coarse estimation is necessary, as those coarse estimates will tell us which are the occupied range bins and velocity bins. Without the coarse estimates, one would need to discretize the entire range or velocity space, which would increase the complexity of the estimation problem. The complexity would be even higher when there are multiple targets, in which case one would  need to try all possible combinations to correctly pair the estimates.

\vspace{-1mm}
\section{Numerical Results}
\vspace{-1mm}
In this section, the performance of the proposed system is evaluated via simulations. The system parameters used are listed in Table~\ref{table:parameters}. During the simulations, the target coefficients were set to $1$. Note again that the coefficients are not required to estimate the target parameters. The binary stream is modulated into data symbols via quadrature phase-shift keying (QPSK), i.e., each data symbol consists of two bits.

\vspace{-0mm}
\begin{table}[t]
\caption{System Parameters}
\vspace{-2mm}
\label{table:system_parameters}
\centering
\begin{tabular}{ ||c||c|c|  }
 \hline
 Parameter & Symbol & Value\\
 \hline
 Center frequency   & $f_c$    &$24$ GHz\\
 Subcarrier spacing &   ${\Delta f} $  & $120$ kHz\\
 Duration of  OFDM symbol & $T_p$ & $8.92\mu$s\\
 Number of subcarriers & $N_s$ & 64\\
 Number of OFDM symbols & $N_p$ & 256\\
  Number of transmit antennas & $N_t$ & 8\\
 Number of radar receive antennas & $N_r$ & 24\\
 Angle of CU & $\theta_c$ & 60\degree \\
 Receive antenna spacing distance & $d_r$ & 0.5$\lambda$\\
 Transmit antenna spacing distance & $d_t$ & 0.5$\lambda$ \\
 \hline
\end{tabular}
\label{table:parameters}
\vspace{-0mm}
\end{table}

\begin{figure}
    \vspace{-5mm}
    \centering
    \includegraphics[width = 6cm]{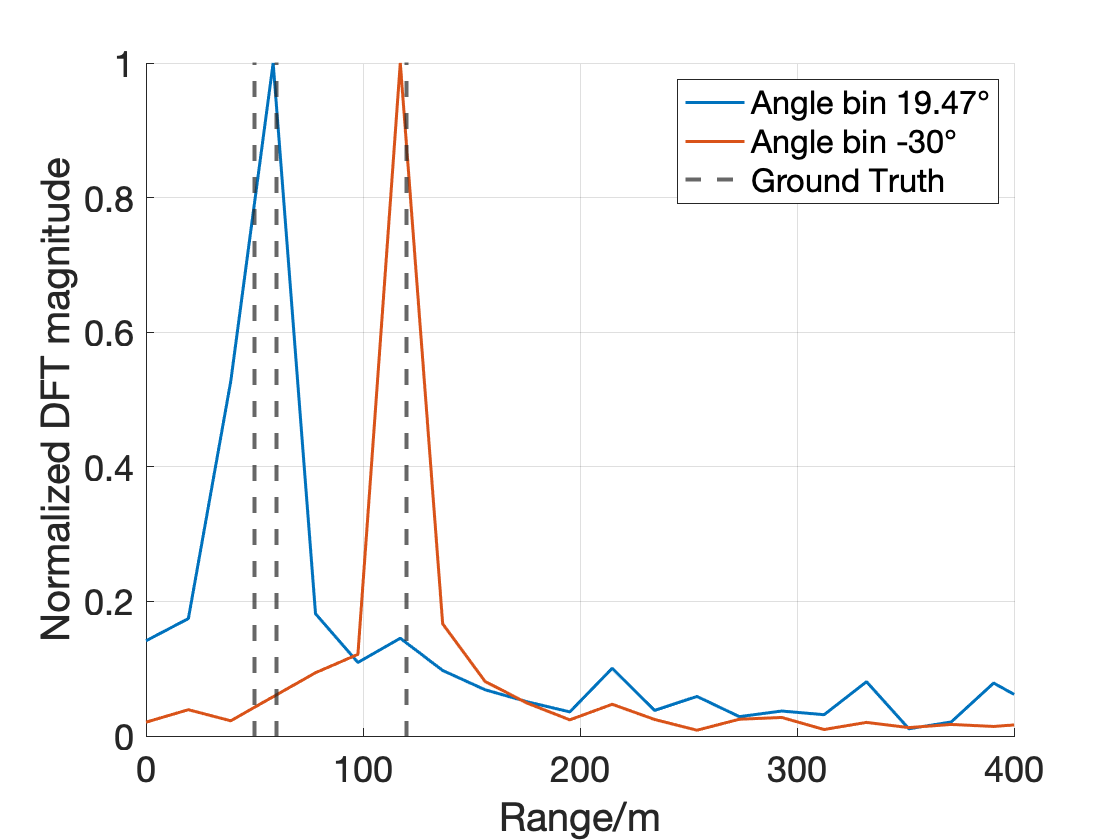}
    \vspace{-4mm}
    \caption{IDFT based range estimation in two angle bins.}
    \label{fig:range}
    \vspace{-7mm}
\end{figure}

\begin{figure}[t]
    \centering
    \includegraphics[width = 6cm]{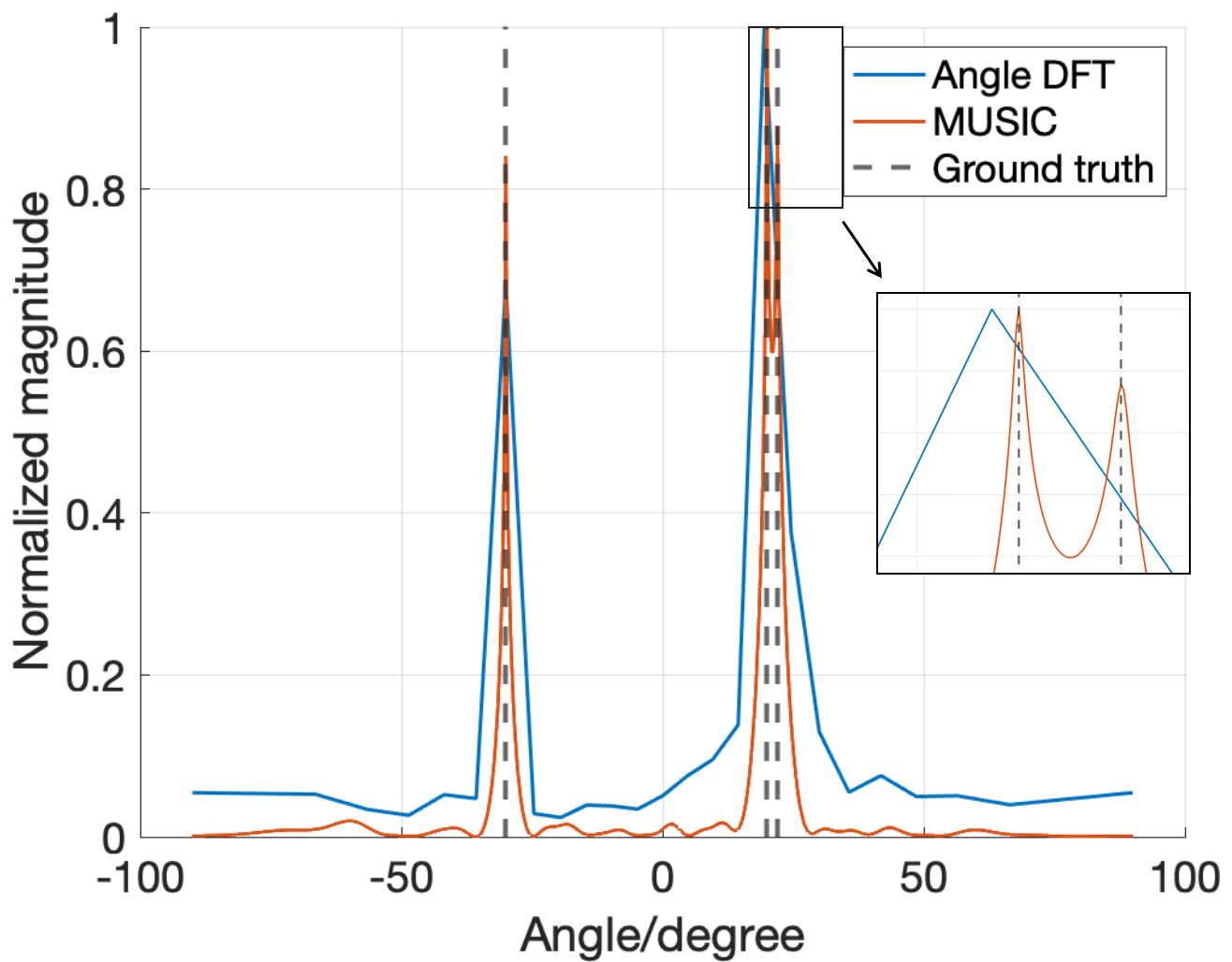}
    \vspace{-4mm}
    \caption{Coarse angle estimation on receive array and refined estimation via the MUSIC algorithm.}
    \label{fig: angle}
    \vspace{-5mm}
\end{figure}

\begin{figure}
    \centering
    \includegraphics[width = 6cm]{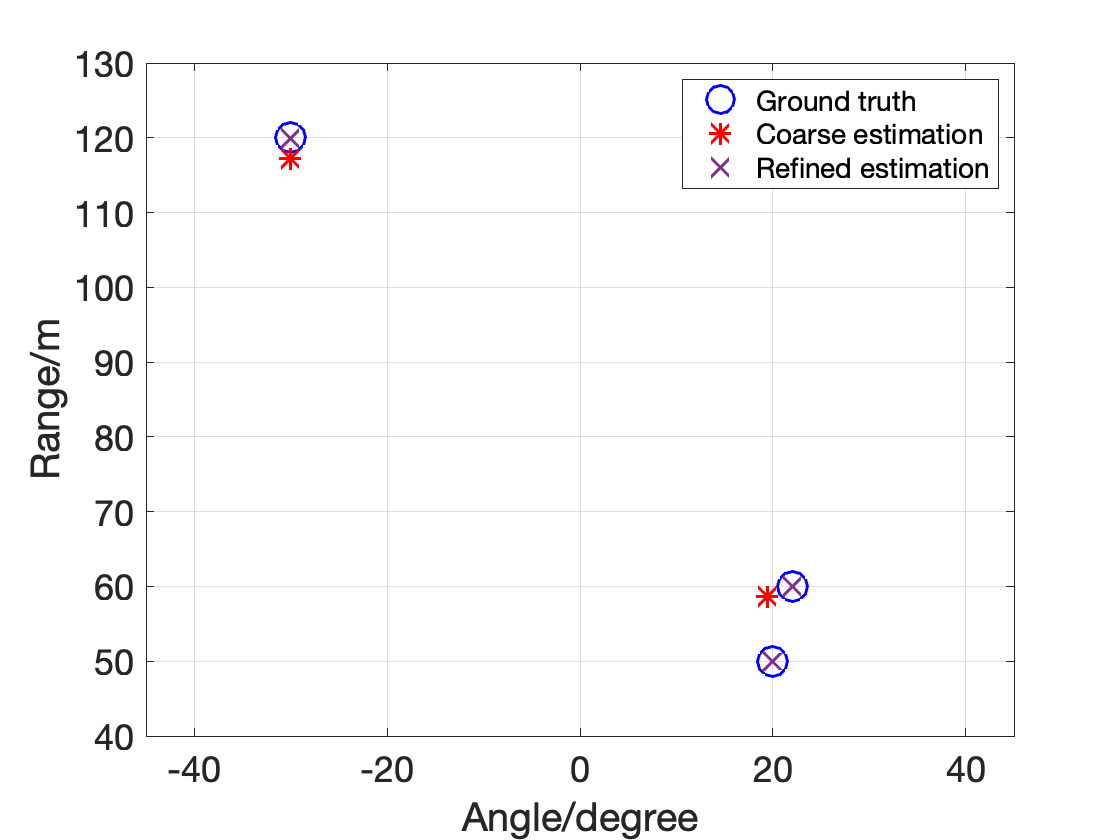}
    \vspace{-3mm}
    \caption{Coarse estimation results and refined estimation results.}
    \vspace{-7mm}
    \label{fig:result}
\end{figure}

The $w_n$ were set to focus the transmitted energy towards the CU and also the data symbols remain in tack in that direction. To enable the DM, $\te = n/N_t$ and $\dt_n = N_t-1/N_t$ for $n = 0,1,\dots, N_t-1$ are used~\cite{Ding2019time_modulated}.
Here we considered $3$ point targets in the far field of the radar array, at angles, ranges, and velocities as shown in Table~\ref{table:radar_result}. 
We should note that based on the aperture of the receive array of $24$ antennas, these targets fall into $2$ different angle bins;  the two targets that fall into the same angle bin have very close ranges but  different speeds.
In the simulation, the SNR was set to $10$dB,

Following the target parameter estimation algorithm described in Sec.~\ref{sec: refine}, we first obtained coarse target angle estimates based on the physical array of $N_r=24$ antennas. The result is shown in Fig.~\ref{fig: angle} in blue.
As one can see, due to the low angle resolution of the receive array,  $2$ targets fell into the same angle bin, i.e., $19.47\degree$ and the other target fell into the angle bin at $-30\degree$.
Subsequently, within each angle bin, target ranges were estimated via \eqref{eq: element-wise division}-\eqref{rangeind}. The results are shown in Fig.~\ref{fig:range}. For the two targets within the same angle bin, their ranges, i.e., $50m$ and $60m$, are too close that their range peak also merged and thus only one peak was seen.
The results are shown in Fig.~\ref{fig:result} in red asterisks, while the ground truth is shown in blue circles. 

Note that, the coarse estimation relies on element-wise division on the scrambled symbols. To examine the potential zero-denominator problem, we repeated $1000$ random simulations, only $0.0074\%$ of symbols were smaller than $0.01$ and none was smaller than $0.001$.

After range estimation, velocities were estimated based on the range peaks following \eqref{veloind}. The coarse target estimation results are listed in the left column of Table~\ref{table:radar_result}. As one can see,  the targets inside the same angle-range bin are separable with respect to their velocities, but have the same angle and range.

\begin{table}[t]
\caption{Target Parameter  Estimation }
\vspace{-2mm}
\begin{tabular}{ |c|c| }
\hline
\multicolumn{2}{|c|}{Target parameters} \\
\hline
\multicolumn{2}{|c|}{ ($20\degree, 50m, -10m/s$), 
($22\degree, 60m, 10m/s$), 
($-30\degree, 120m, 20m/s$)}\\
 \hline
 Coarse estimation & Refined estimation \\
\hline
 $(19.47\degree, 58.59$m, $-9.38m/s$) & $(20\degree, 50.00$m, $-10.08m/s$)  \\ 
 $(19.47\degree, 58.59$m, $9.38m/s$) & $(22\degree, 60.16$m, $10.08m/s$)\\ 
 ($-30\degree, 117.19$m, $21.09m/s$) & ($-30\degree, 120.12$m, $19.92m/s$) \\ 
\hline
\end{tabular}
\label{table:radar_result}
\vspace{-5mm}
\end{table}

To refine the coarse target estimates, we followed Algorithm~\ref{alg:angle-range}.
We applied the MUSIC algorithm to refine the angle estimates, and the results are plotted in Fig.~\ref{fig: angle} in red. As one can see, the two targets that fell in the same angle bin are now successfully resolved and have two clear peaks. Based on the refined angles, we formulated a least-squares problem (see \eqref{eq: MLE}). In the simulation, we discretized the range space within the occupied angle-range bin on $101$ grid points. Solving the formulated least-squares problem we obtained the results shown in Fig.~\ref{fig:result} in purple cross. This time, the two closely spaced targets are also resolved in range. 

In order to refine the velocity estimates, we formulated another least-squares problem using the refined angle-range estimates. However, due to the large size of the formulated problem, we discretized the velocity space into $11$ grid points to reduce the computation overhead.
The refined target estimates are shown in the right column of Table~\ref{table:radar_result}. Clearly, the refined target estimates match well with the ground truth.

\vspace{-1mm}
\section{Conclusion}
\vspace{-1mm}
In this paper, we have proposed to realize the sensing capability based on a communication OFDM DM transmitter.
Through the proposed novel estimation method, we have addressed the coupling between the scrambled symbols and target parameters and the target parameters can be coarsely estimated. To refine those coarse estimates, we have proposed an estimation refinement algorithm where the angles are first refined via the MUSIC algorithm and then the ranges are refined via a least-squares approach based on the refined angles. The target velocities can also be refined in a similar way. Although the coarse radar estimation is limited by the size of receive array and the number of subcarriers, the proposed estimation refinement algorithm is able to provide estimation results with high resolution at a cost of higher complexity.

\vspace{-2mm}
\bibliographystyle{IEEEtran}
\bibliography{ref.bib}

\end{document}